\PassOptionsToPackage{unicode}{hyperref}
\PassOptionsToPackage{hyphens}{url}
\PassOptionsToPackage{dvipsnames,svgnames,x11names}{xcolor}
\documentclass[
  12pt]{article}

\usepackage{amsmath,amssymb}
\usepackage{iftex}
\ifPDFTeX
  \usepackage[T1]{fontenc}
  \usepackage[utf8]{inputenc}
  \usepackage{textcomp} 
\else 
  \usepackage{unicode-math}
  \defaultfontfeatures{Scale=MatchLowercase}
  \defaultfontfeatures[\rmfamily]{Ligatures=TeX,Scale=1}
\fi
\usepackage{lmodern}
\ifPDFTeX\else  
\fi
\IfFileExists{upquote.sty}{\usepackage{upquote}}{}
\IfFileExists{microtype.sty}{
  \usepackage[]{microtype}
  \UseMicrotypeSet[protrusion]{basicmath} 
}{}
\makeatletter
\@ifundefined{KOMAClassName}{
  \IfFileExists{parskip.sty}{%
    \usepackage{parskip}
  }{
    \setlength{\parindent}{0pt}
    \setlength{\parskip}{6pt plus 2pt minus 1pt}}
}{
  \KOMAoptions{parskip=half}}
\makeatother
\usepackage{xcolor}
\makeatletter
\ifx\paragraph\undefined\else
  \let\oldparagraph\paragraph
  \renewcommand{\paragraph}{
    \@ifstar
      \xxxParagraphStar
      \xxxParagraphNoStar
  }
  \newcommand{\xxxParagraphStar}[1]{\oldparagraph*{#1}\mbox{}}
  \newcommand{\xxxParagraphNoStar}[1]{\oldparagraph{#1}\mbox{}}
\fi
\ifx\subparagraph\undefined\else
  \let\oldsubparagraph\subparagraph
  \renewcommand{\subparagraph}{
    \@ifstar
      \xxxSubParagraphStar
      \xxxSubParagraphNoStar
  }
  \newcommand{\xxxSubParagraphStar}[1]{\oldsubparagraph*{#1}\mbox{}}
  \newcommand{\xxxSubParagraphNoStar}[1]{\oldsubparagraph{#1}\mbox{}}
\fi
\makeatother

\usepackage{longtable,booktabs,array}
\usepackage{calc} 
\usepackage{etoolbox}
\makeatletter
\patchcmd\longtable{\par}{\if@noskipsec\mbox{}\fi\par}{}{}
\makeatother
\IfFileExists{footnotehyper.sty}{\usepackage{footnotehyper}}{\usepackage{footnote}}
\makesavenoteenv{longtable}
\usepackage{graphicx}
\makeatletter
\def\maxwidth{\ifdim\Gin@nat@width>\linewidth\linewidth\else\Gin@nat@width\fi}
\def\maxheight{\ifdim\Gin@nat@height>\textheight\textheight\else\Gin@nat@height\fi}
\makeatother
\setkeys{Gin}{width=\maxwidth,height=\maxheight,keepaspectratio}
\makeatletter
\def\fps@figure{htbp}
\makeatother

\makeatletter
\@ifpackageloaded{caption}{}{\usepackage{caption}}
\AtBeginDocument{%
\ifdefined\contentsname
  \renewcommand*\contentsname{Table of contents}
\else
  \newcommand\contentsname{Table of contents}
\fi
\ifdefined\listfigurename
  \renewcommand*\listfigurename{List of Figures}
\else
  \newcommand\listfigurename{List of Figures}
\fi
\ifdefined\listtablename
  \renewcommand*\listtablename{List of Tables}
\else
  \newcommand\listtablename{List of Tables}
\fi
\ifdefined\figurename
  \renewcommand*\figurename{Figure}
\else
  \newcommand\figurename{Figure}
\fi
\ifdefined\tablename
  \renewcommand*\tablename{Table}
\else
  \newcommand\tablename{Table}
\fi
}
\@ifpackageloaded{float}{}{\usepackage{float}}
\floatstyle{ruled}
\@ifundefined{c@chapter}{\newfloat{codelisting}{h}{lop}}{\newfloat{codelisting}{h}{lop}[chapter]}
\floatname{codelisting}{Listing}

\makeatother
\makeatletter
\@ifpackageloaded{caption}{}{\usepackage{caption}}
\@ifpackageloaded{subcaption}{}{\usepackage{subcaption}}
\makeatother

\ifLuaTeX
  \usepackage{selnolig}  
\fi
\usepackage[]{natbib}
\usepackage{bookmark}

\IfFileExists{xurl.sty}{\usepackage{xurl}}{} 
\hypersetup{
  pdftitle={Title},
  pdfauthor={Author 1; Author 2},
  pdfkeywords={3 to 6 keywords, that do not appear in the title},
  colorlinks=true,
  linkcolor={blue},
  filecolor={Maroon},
  citecolor={Blue},
  urlcolor={Blue},
  pdfcreator={LaTeX via pandoc}}

\usepackage{amsfonts}
\usepackage{amsthm}%
\usepackage{mathrsfs}%
\usepackage{manyfoot}%
\usepackage{algorithm}%
\usepackage{algorithmicx}%
\usepackage{algpseudocode}%
\usepackage{listings}%
\usepackage{dsfont}
\usepackage{bm}
\usepackage{physics}

\newcommand{\prob}{\operatorname{P}}
\newcommand{\E}{\operatorname{E}}
\newcommand{\diff}{\mathrm{d}}
\newcommand{\bX}{\bm X}
\newcommand{\bY}{\bm Y}
\newcommand{\bZ}{\bm Z}

\newcommand{\hX}{\widehat{X}}

\newcommand{\1}{\operatorname{\mathds{1}}}
\newcommand{\Vq}{\mathcal{V}^q}
\newcommand{\cbr}[1]{\left\{ {#1} \right\}}
\newcommand{\rbr}[1]{\left( {#1} \right)}

\newcommand{\reals}{\mathbb{R}}

\theoremstyle{thmstyleone}%
\newtheorem{theorem}{Theorem}%
\theoremstyle{thmstyletwo}%
\theoremstyle{thmstylethree}%

\newcommand{\anon}{1}

\begin{document}

\def\spacingset#1{\renewcommand{\baselinestretch}%
{#1}\small\normalsize} \spacingset{1}


\if1\anon
{
  \title{\bf Transformed Linear Prediction for Extremes}
  \author{Jeongjin Lee\thanks{
    Corresponding author. Email: j.lee58@lancaster.ac.uk
    \href{https://orcid.org/0000-0001-5655-3358}{ORCID: 0000-0001-5655-3358}
    }\hspace{.2cm}\\
    School of Mathematical Sciences, Lancaster University,\\ Fylde College, Lancaster, LA1 4YF, United Kingdom\\
    and \\
    Daniel Cooley \\
    Department of Statistics, Colorado State University,\\ Fort Collins, 80523, Colorado, United States of America}
  \maketitle
} \fi

\if0\anon
{
  \bigskip
  \bigskip
  \bigskip
  \begin{center}
    {\LARGE\bf Transformed Linear Prediction for Extremes}
\end{center}
  \medskip
} \fi

\bigskip
\begin{abstract}
We address the problem of 
prediction for extreme observations by proposing an extremal linear prediction method.
We construct an inner product space of nonnegative random variables derived from transformed-linear combinations of independent regularly varying random variables.
Under a reasonable modeling assumption, the matrix of inner products corresponds to the tail pairwise dependence matrix, which can be easily estimated.
We derive the optimal transformed-linear predictor via the projection theorem, which yields a predictor with the same form as the best linear unbiased predictor in non-extreme settings.
We quantify uncertainty for prediction errors by constructing prediction intervals based on the geometry of regular variation.
We demonstrate the effectiveness of our method through a simulation study and its applications to predicting high pollution levels, and extreme precipitation.
\end{abstract}

\noindent%
\vfill

\newpage
\spacingset{1.8} 

\section{Introduction}
\label{sec:intro}

\begin{figure}
\centerline{\includegraphics[width=2.4in,height=\textheight]{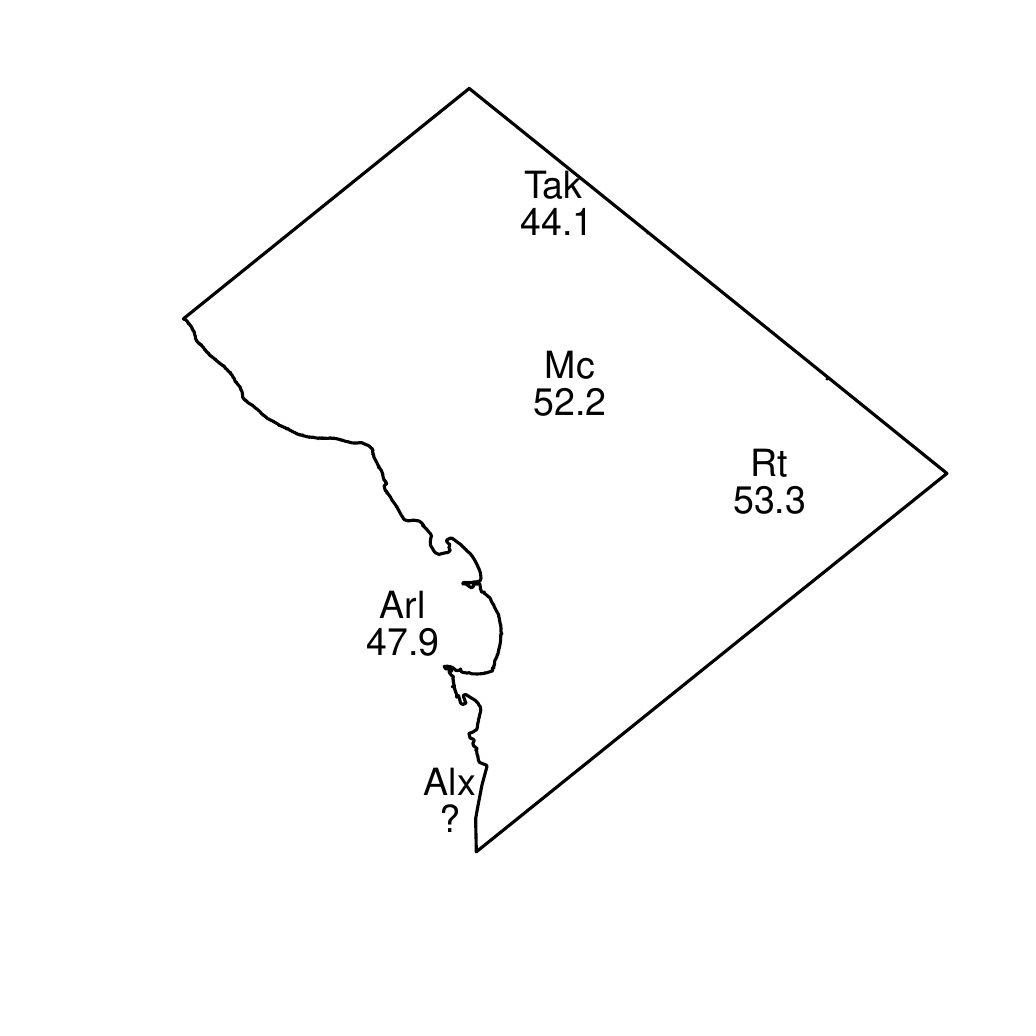}}
\vspace{-1cm}
  \caption{Maximum NO$_2$ measurements for January 23, 2020.  All observations are above the empirical .98 quantile for each location.}
  \label{fig: washingtonDC}
\end{figure}

Prediction (or imputation or interpolation) of unobserved quantities is a key objective in statistical analysis.
Figure \ref{fig: washingtonDC} shows the one-hour maximum measurements of the air pollutant nitrogen dioxide (NO$_2$) at four monitoring stations in the Washington DC area on January 23, 2020.
The measurements, given in parts per billion, are at notably high levels; each measurement exceeds its station's empirical 0.98 quantile for the year, and the Arlington station (Arl) records its highest level of the year.
Given these measurements, a natural question is to predict the NO$_2$ level at a nearby unmonitored location such as Alexandria VA, marked ``Alx'' in Figure \ref{fig: washingtonDC}, which had NO$_2$ monitoring prior to 2015.


In a classical statistical setting, if the joint distribution of all variates were known, the conditional distribution would provide complete information about the variate of interest given the observed values.
However, the distribution of the air pollution data is unknown, clearly non-Gaussian, and lacks a clear candidate for a joint distribution.
Furthermore, extreme value analysis would caution against using a model that had been fit to the entire dataset to characterize behavior in the (joint) tail.

In this work, we propose a novel linear prediction method specifically for extremes.
Linear methods are appealing as they offer a straightforward approach to prediction by applying weights to each observation.
In non-extreme settings, linear prediction methods utilize information in the covariance matrix, which does not require specification of the full joint distribution, and which is easily estimated using only pairwise relationships even when the dimension is high.
Linear prediction methods provide the best linear unbiased prediction (BLUP) weights in terms of mean squared prediction error (MSPE). 
Prediction uncertainty is typically characterized via MSPE, which is directly obtainable from the covariance matrix.
Linear methods are widely applied across various domains, including spatial statistics (e.g., kriging), time series forecasting, and multivariate analysis.

Our proposed extremal prediction method is similar in spirit to classical linear prediction but is designed exclusively for extreme values.
To establish a framework for modeling dependence in the upper tail, we rely on regular variation on the positive orthant, as summarized in Section~\ref{sec:regularvariation}. 
After a polar coordinate transformation, the angular measure fully characterizes tail dependence in the limit of a regularly varying random vector.
However, fully estimating the angular measure is challenging even in moderate dimensional settings.
Rather than fully characterizing the angular measure, we summarize pairwise tail dependencies in the tail pairwise dependence matrix, which has properties analogous to a covariance matrix, as described in Section~\ref{sec:TPDM}.

Defining our extremal linear predictor requires a vector space, which we develop in Section \ref{sec:ipspacePred}.
Because we have interest in only one direction (e.g.; we wish to predict when pollution levels are high), we model only in the positive orthant.
We therefore construct a vector space of {\em non-negative} random vectors derived from transformed-linear combinations of independent regularly varying random variables. 
We then define an inner product and, via the projection theorem, derive the best transformed-linear predictor whose form is analogous to the BLUP in non-extreme linear prediction.


We also develop a method for quantifying uncertainty in our prediction errors, as detailed in Section~\ref{sec:UQ}.
Prediction intervals are based on the polar geometry of regular variation rather than the elliptical geometry underlying standard linear prediction and leading to the use of MSPE.
The use of polar geometry is crucial because, in this framework, the error magnitude is conditional on the size of the predicted value.
The effectiveness of our method is demonstrated through a simulation study in Section~\ref{sec:simulation} and its application to the Washington air pollution data and a higher-dimensional precipitation data set in Section~\ref{sec:applications}.

\section{Background}
\label{sec:background}


\subsection{Regular variation on the positive orthant}
\label{sec:regularvariation}

Multivariate regular variation on the positive orthant is closely linked to classical extreme value analysis \citep[Chapter 6]{deHaan2007} and the positive orthant is a natural support for multivariate exceedances. 
\cite{resnick2007} gives a comprehensive treatment of regular variation.
Let $\bm{X}$ be a $p$-dimensional random vector that takes values in $\mathbb{R}_{+}^{p}=[0,\infty)^p$.
The random vector $\bm{X}$ is regularly varying (denoted $RV_+^p(\alpha)$) if there exists a function $b(s) \rightarrow \infty$ as $s \rightarrow \infty$ and a non-degenerate limit measure $\nu_{\bm X}$ 
such that 
\begin{equation}
\label{eq: regVar1}
  s\prob(b(s)^{-1}\bm{X}\in \cdot)\xrightarrow{v} \nu_{\bm{X}}(\cdot),
\end{equation}
where $\xrightarrow{v}$ indicates vague convergence in $M_+(\mathbb{E})$, the space of nontrivial Radon measures on the Borel subsets of $\mathbb{E}:=[0,\infty]^{p} \setminus \{\bm{0}\}$.
The normalizing function $b(s)$ is regularly varying with an index of $1/\alpha$; that is, it takes the form $b(s)=L(s) s^{1/\alpha}$, where $L(s)$ is a slowly varying function at infinity, i.e., $\lim_{s\rightarrow\infty}L(sx)/L(s)=1$.
For any set $C \subset \mathbb{E}$ and $k > 0$, the limit measure has the scaling property $\nu_{\bm{X}}(kC)=k^{-\alpha}\nu_{\bm{X}}(C)$, and tail index $\alpha > 0$ determines the tail heaviness.

The scaling property implies regular variation can be more easily understood in a polar geometry.
Given any norm $\|\cdot\|$ on $\reals^p$, define the unit sphere restricted to the positive orthant as $\Theta_{+}^{p-1}=\{\bm{x}\in \reals^{p}:\|\bm{x}\|=1\}\cap \mathbb{E}$.
For $r>0$ and Borel set $B\subset \Theta_{+}^{p-1}$, define the set $C(r,B)=\{\bm{x}\in [0,\infty)^p\setminus{\{\bm 0\}}:\|\bm{x}\|>r, \bm{x}/\|\bm{x}\| \in B\}$.
The scaling property implies $\nu_{\bm{X}}{(C(r,B))}=r^{-\alpha}\nu_{\bX}(C(1, B))$, so defining $\bX$'s angular measure $H_{\bX}(B) := \nu_{\bX}(C(1,B))$, we obtain an equivalent polar-coordinate formulation of regular variation
\begin{equation*}
    s\prob\rbr{\rbr{b(s)^{-1}\|\bX\|,\bX/\|\bX\|}\in \cdot}\xrightarrow{v} \nu_{\alpha}\times H_{\bX}
\end{equation*}
in $M_+((0,\infty]\times \Theta_+^{p-1})$ \citep[Theorem 6.1,][]{resnick2007}, where $\nu_{\alpha}(r,\infty]=r^{-\alpha}$.
The measure's intensity function in terms of polar coordinates is 
\begin{equation}
    \label{eq:nu}
    \nu_{\bm{X}}(\diff r\times \diff\bm w)=\alpha r^{-\alpha-1}\diff r\diff H_{\bm{X}}(\bm w).
\end{equation}
As the angular measure $H_{\bm X}$ fully describes tail dependence in the limit, it is of primary importance in the modeling of multivariate extremes.

The normalizing function $b(s)$ can be any function which results in a nontrivial limit in (\ref{eq: regVar1}).
Once $b(s)$ is chosen, we are able to compare the relative magnitudes of the components of $\bX$.
Letting $C_i(r) = \{\bm x \in \mathbb{R}^p_+ \mid x_i > r\}$ for $i \in \{1, \ldots, p\}$, 
\begin{equation}
  \label{eq:scale}
  \lim_{s \rightarrow \infty} s \prob ( b(s)^{-1} \bm X \in C_i(r) ) 
  = r^{-\alpha} \lim_{s \rightarrow \infty} s \prob ( b(s)^{-1} X_i > 1) := r^{-\alpha} \kappa_i.
\end{equation}
We will refer to $\kappa_i$ as the `scale' of $X_i$.   
The scale is also contained in the angular measure as 
\begin{equation}
  \label{eq:scaleH}
  \kappa_i = \int_{\Theta_+^{p - 1}} \int_{1/w_i}^\infty \alpha r^{-\alpha - 1} \diff r \diff H_{\bX}(\bm w) = \int_{\Theta_+^{p - 1}} w_i^\alpha \diff H_{\bX}(\bm w).
\end{equation}
Additionally, the overall mass of the angular measure is determined by the choice of $b(s)$, as
$$
  H_{\bX}(\Theta_+^{p-1}) = \lim_{s \rightarrow \infty} \prob(b(s)^{-1} \|\bX\| > 1) := m.
$$
Often $b(s)$ is chosen such that that $\prob(\| \bX \| > b(s) ) = s^{-1}$ resulting in $H_{\bX}$ being a probability measure, but we make no such specification.
When we discuss linear constructions of regularly varying random vectors in Section \ref{sec:linearConstructions}, we will choose $b(s)$ such that the noise term has scale 1.

\subsection{Tail pairwise dependence matrix}
\label{sec:TPDM}

For a general $\bm X \in RV_+^p(\alpha)$, if $p$ is even moderately large, it is challenging to model or estimate the angular measure $H_{\bm X}$.
Rather than fully characterize $H_{\bm X}$, we will summarize tail dependence via a matrix of pairwise summary measures, the tail pairwise dependence matrix (TPDM).
Let $\Sigma_{\bm X} =\{ \sigma_{ij}\}_{i,j\in \{1, \ldots p\}}$ denote the $p \times p$ TPDM where 
\begin{equation}
\label{eq: TPDM}
  \sigma_{ij}=\sigma(X_i, X_j) =\int_{\Theta_{+}^{p-1}} {w_{i}^{\alpha/2}w_{j}^{\alpha/2}} \diff H_{\bm{X}}(\bm w).
\end{equation}
The TPDM given by (\ref{eq: TPDM}) corresponds to the definition given by \cite{kiriliouk2022}.
\cite{cooley2019decompositions} define the TPDM's elements as $\sigma_{ij}=\int_{\Theta_{+}^{p-1}} {w_{i}w_{j}} \diff H_{\bm{X}}(\bm w)$, but focus on the case when $\alpha = 2$ where the definitions are equivalent.

For prediction, the TPDM will play the role of the covariance matrix and it shares similar properties.
First, $\Sigma_{\bm X}$ is positive semi-definite; see Section 4 of \cite{cooley2019decompositions}.
Similar to how the diagonal elements of a covariance matrix give the variance of a random vector's elements, comparing (\ref{eq:scaleH}) to (\ref{eq: TPDM}) shows that the TPDM's diagonal element $\sigma_{ii}$ gives the scale of $X_i$.
Moreover, if $\sigma_{ij} = 0$, then $H_{\bm X}(\{\bm w \in \Theta_+^{p-1} \mid w_i > 0, w_j > 0\}) = 0$, implying that $X_i$ and $X_j$ are asymptotically independent \citep[][Section 6.5.1]{resnick2007}.


\subsection{Linear constructions for multivariate regular variation}
\label{sec:linearConstructions}
Defining random variables as linear combinations of independent `noise' variables (i.e., $X_i = \bm a_i^\top \bm Z$, with $\bm a_i \in \mathbb{R}^q$ and $\bZ$ a $q$-dimensional vector of i.i.d. variables), is a simple way to construct a collection of random variables with nontrivial dependence.
Below we review similar constructions where the noise terms are regularly varying, eventually describing transformed-linear construction which will be used to construct our vector space in Section \ref{sec:innerProductSpace}.

\subsubsection{Standard linear and max linear constructions}
Let $b(s)$ be given and let $\bZ = (Z_1, \ldots, Z_q)^\top \in RV_+^q(\alpha)$ be a vector of independent nonnegative regularly varying random variables with unit scale; i.e., $P(Z_1 > b(s)) = s^{-1}$.
For $A \in \mathbb{R}^{p \times q}$, consider the standard linear construction $\bX = A\bZ$, and denote its $i$th element as $X_i = \bm a_{i}^\top \bm Z$, where $\bm a_{i}^\top$ is $A$'s $i$th row.
For $X_i$ to be non-trivial, we will assume that $\bm a_{i} \neq \bm 0$ for all $i = 1, \ldots, p$.
The random vector $\bX = A\bZ$ is jointly regularly varying, and if defined using the same $b(s)$, its angular measure is
$$
    H_{A\bZ}(\cdot) = \sum_{j = 1}^q \|  \bm a_{\cdot j} \|^\alpha \delta_{ \bm a_{\cdot j} / \|  \bm a_{\cdot j} \|}(\cdot),
$$
where $\bm{a}_{\cdot j}$ is the $j$th column of $A$, and $\delta$ is the Dirac mass function (see Section 1.1 of the supplement).
Thus, the angular measure consists of $q$ point masses.
Since $A$ has not (yet) been restricted to nonnegative values, $\bX$ is not confined to the positive orthant.
Consequently, definition (\ref{eq: regVar1}) would be defined on Borel subsets of $[-\infty, \infty] \setminus \{\bm 0 \}$, and the angular measure would exist on $\Theta^{p-1} = \{\bm x \in \mathbb{R}^p : \| \bm x \| = 1\}$ rather than $\Theta_+^{p-1}$; see Section 6.5.5 of \cite{resnick2007} for a more comprehensive discussion of regular variation beyond the positive orthant.
The scale of $X_i$, now defined as $\kappa_i = \lim_{s \rightarrow \infty} s \prob ( b(s)^{-1} | X_i | > 1)$, is $\sum_{j = 1}^q | a_{ij} |^\alpha.$
As our intention is to model only on the positive orthant, one could consider constraining $A$ to consist only of nonnegative elements.
However, such a construction would not be amenable to a vector space structure, as the additive inverse of $X_i$, given by $-\bm a_{i}^\top \bZ$, would not meet the nonnegative restriction.

Max-linear constructions replace the addition operation in a linear combination with the maximum operation: given $A \in \mathbb{R}^{p \times q}$, $\bm a_{i}$, and $\bZ \in RV^q_+(\alpha)$ as above, define $\bX = A\otimes_{\max}\bZ$, where the $i$th element is $X_i = \bm a_{i}^\top \otimes_{\max} \bZ = \max_{j = 1, \ldots q} a_{ij} Z_j$. 
Note that with a max-stable construction, it is not required that $A$'s coefficients be non-negative for $\bX$ to have support on the positive orthant.
For $x > 0$, and so long as $\max_{j = 1, \ldots, q} a_{ij} > 0$, $\prob(X_i > x) = \prob(\max_{j=1,\ldots,q} a_{ij} Z_j > x) = \prob(\max_{j=1,\ldots,q} a_{ij}^{(0)} Z_j > x),$ where $a^{(0)} = \max(a,0)$.
$X_i$ is regularly varying and $\lim_{s \rightarrow \infty}s\prob(b(s)^{-1} X_i > 1) = \sum_{j = 1}^q (a_{ij}^{(0)})^\alpha \lim_{s \rightarrow \infty}s\prob(b(s)^{-1} Z_1 > 1)$, so $\kappa_i = \sum_{j = 1}^q (a_{ij}^{(0)})^\alpha$ \citep[][Lemma 3.11]{jessen2006}.
Assuming $\max_{j = 1, \ldots q} a_{ij} > 0$ for all $i$ and $\max_{i = 1, \ldots p} a_{ij} > 0$ for all $j$, the random vector $\bX = A\otimes_{\max}\bZ$ is regularly varying, and if normalized by the same $b(s)$ as $Z_j$, then
\begin{equation}
  \label{eq:linAngMsr}
    H_{A\otimes_{\max}\bZ}(\cdot) = \sum_{j = 1}^q \|  \bm a_{\cdot j}^{(0)} \|^\alpha \delta_{ \bm a_{\cdot j}^{(0)} / \|  \bm a_{\cdot j}^{(0)} \|}(\cdot),
\end{equation}
where the zero operation is applied componentwise.
However, introducing negative coefficients into $A$ does not allow the max-linear construction to solve the problem of the additive inverse needed for a vector space.

Max-linear constructions have a long history in the modeling of extremes (e.g., \cite{schlather2002inequalities}; \cite{FOUGERES2013109}) as they are connected to the idea of max-stability of classical extreme value theory.
Specifically, if $Z_i$ is Fr\'echet($\alpha$); i.e., $\prob(Z_i \leq z) = \exp(-z^{-\alpha})$, then $X_i$ is max-stable and $\prob(X_i \leq x) = \exp[-(\sum_{j = 1}^q a_{ij}^\alpha x^{-\alpha})]$ if $a_{ij} \geq 0$ for all $j = 1, \ldots q$.
Recently, max-linear constructions have been used to model directed acyclic graphs for extreme phenomena \citep{kluppelberg2021estimating,gissibl2018max}.

\subsubsection{Transformed linear constructions}
\label{sec:transformedOperations}

\cite{cooley2019decompositions} defined transformed linear operations in order to perform `linear' operations for vectors in the target space of the positive orthant.
Consider $\bm x \in \mathbb{R}_{+}^{p}=[0,\infty)^{p}$ and let $t$ be a monotone bijection mapping from $\mathbb{R}$ to $\mathbb{R}_{+}$, with $t^{-1}$ its inverse.
For $\bm y \in \mathbb{R}^{p}$, $t(\bm y)$ applies the transform componentwise.
For $\bm{x}_1, \bm{x}_2 \in \mathbb{R}_{+}^{p}$, we define vector addition as $\bm{x}_1 \oplus \bm{x}_2 = t\{t^{-1}(\bm{x}_1)+t^{-1}(\bm{x}_2)\}$, scalar multiplication as $a\circ \bm{x}_1=t\{at^{-1}(\bm{x}_1)\}$ for $a\in \mathbb{R}$, the additive identity as $\bm{0}:=t(\bm{0})$, the additive inverse of $\bm{x}_1$ as $\ominus \bm{x}_1 = t\{-t^{-1}(\bm{x}_1)\}$, and the subtraction as $\bm{x}_1\ominus \bm{x}_2=t\{t^{-1}(\bm{x}_1)-t^{-1}(\bm{x}_2)\}$.
It is straightforward to show that $\mathbb{R}_{+}^{p}$ with these transformed-linear operations is a vector space as it is isomorphic to $\mathbb{R}^{p}$ with standard operations.

We now consider applying transformed linear operations to non-negative regularly varying random vectors $\bm X_k$ as in (\ref{eq: regVar1}) with limit measure $\nu_{\bm X_k}(\cdot)$, $k=1,2$.
We need two conditions to preserve regular variation in the upper tail.
First, the function $t$ is chosen such that 
\begin{equation}
\label{eq:transform}
    \lim\limits_{y\to\infty} t(y)/y=\lim\limits_{x\to\infty} t^{-1}(x)/x=1,
\end{equation}
ensuring that $t$ negligibly affects large values. 
Further we assume that the marginals meet a {\em lower tail condition}, which ensures that when $\bX_{k}$ has mass near zero and the coefficient $a$ is negative, the term $a \circ \bm X_k$ does not affect the upper tail.
Denoting the $i$th element of $\bm X_k$ as $X_{ki}$, the lower tail condition requires $\prob(X_{ki} < x) \rightarrow 0$ sufficiently fast as $x \rightarrow 0$ so that for any $a < 0$, $\lim_{s \rightarrow \infty} sP(b(s)^{-1} a \circ X_{ki} > 1) = 0$.
To make things specific, \cite{cooley2019decompositions} consider $t(y)=\log\cbr{1+\exp(y)}$ with its inverse $t^{-1}(x)=\log\cbr{\exp(x)-1}$, and assume the lower tail condition 
\begin{equation}
  \label{eq:lowertail}
    s\prob\{X_{ki} \le \exp(a b(s))\}\to 0, \quad\mbox{ as } s\rightarrow\infty,
\end{equation}
$i= 1, \ldots, p$, for all $a < 0$.
Applying transformed linear operations, \cite{cooley2019decompositions} show that if $\bm X_1, \bm X_2 \in RV_+^p(
\alpha)$ are independent, 
\begin{eqnarray}
\label{eq:transPlus}
s\prob(b(s)^{-1}(\bm X_1 \oplus \bm X_2)\in \cdot) &\xrightarrow{v}& \nu_{\bm X_1}(\cdot)+\nu_{\bm X_2}(\cdot)\mbox{; and}\\
\label{eq:transMult}
s\prob(b(s)^{-1}(a\circ \bm X_1)\in \cdot) &\xrightarrow{v}&
   \left \{
     \begin{array}{l}
       a^{\alpha}\nu_{\bm X_1}(\cdot) \mbox{ if } a>0\mbox{, and}\\
       0 \,\,\qquad\quad\mbox{ if } a \le 0.
     \end{array}
   \right. 
\end{eqnarray}
Other transforms with the same limiting properties and with appropriately adjusted lower tail condition could be used in place of $t$.

Both properties~\eqref{eq:transPlus} and~\eqref{eq:transMult} together allow for constructing a regularly varying random vector via transformed linear combinations of independent regularly varying random variables.  
As above, let $A \in \mathbb{R}^{p \times q}$ and assume $\max_{i =  1, \ldots, p}a_{ij}>0$ for all $j= 1, \ldots, q$, $\max_{j= 1, \ldots, q}a_{ij}>0$ for all $i= 1, \ldots, p$, and $\bm Z = (Z_1, \ldots Z_q)^\top\in RV_+^q(\alpha)$ be a vector of independent regularly varying random variables with unit scale when normalized by $b(s)$ and which also meet~\eqref{eq:lowertail}.  
Consider the random vector
\begin{equation}
\label{eq:linConst}
    A \circ \bm Z = t(A t^{-1}(\bm Z)) \in RV^p_+(\alpha). 
\end{equation}
$A \circ \bZ$ is similar in behavior to the max-linear construction $A \otimes_{\max} \bm Z$, and in fact shares the same angular measure (\ref{eq:linAngMsr}), see Corollary 1 of \cite{cooley2019decompositions}.
One difference is that if $X_i = \bm a_{i}^\top \circ \bm Z$, then the additive inverse $-X_i = -\bm a_{i}^\top \circ \bZ$ also lies in the positive orthant. 

From (\ref{eq:linAngMsr}), the elements of the TPDM of $A \circ \bZ$ are
$$
  \sigma_{ik} = \sum_{j = 1}^q 
  \left(
    \frac{a_{ij}^{(0)}}{\|\bm a_{\cdot j}^{(0)}\|}
  \right) ^{\alpha/2}
  \left(
    \frac{a_{kj}^{(0)}}{\| \bm a_{\cdot j}^{(0)}\|}
  \right) ^{\alpha/2} 
  \|\bm a_{\cdot j}^{(0)}\|^\alpha
  = \sum_{j = 1}^q
    (a_{ij}^{(0)})^{\alpha/2}
    (a_{kj}^{(0)})^{\alpha/2}.
$$
We note that the TPDM of $A\circ \bZ$ is linear only if $\alpha = 2$, and linearity of the TPDM will be critical to connect to the notion of inner product.
In fact for general jointly regularly varying $(X_1, X_2, X_3)$ (not arising from any particular construction method), $\sigma(a_1 X_1 + a_2 X_2, X_3) = a_1 \sigma(X_1, X_3) + a_2 \sigma(X_2, X_3)$ only if $\alpha = 2$ as shown in Section 1.2 of the supplement.

All of the linearly-constructed models result in angular measures consisting of $q$ point masses.
In theory, these construction methods are rich in the sense that any angular measure can be approximated arbitrarily close by allowing $q$ to increase \citep{FOUGERES2013109, cooley2019decompositions}.
In Section \ref{sec:innerProductSpace}, we will use the transformed-linear construction method to construct a vector space.
For prediction in Section \ref{sec:transLinearPred}, the random vector $\bm Z$ and its dimension $q$ nicely can be considered {\em latent} and are not needed for inference.

\section{Inner product space and linear prediction}
\label{sec:ipspacePred}


We aim to define a vector space of random variables that aligns with the framework of regular variation.
It turns out this is challenging.
For context, consider 
the collection of random variables which share a common normalizing function $b(s)$.
Even with a common $b(s)$, simple properties like closure can become problematic:
if $X_1 \in RV_+^1(\alpha)$ with normalizing function $b(s)$ and $X_2 = X_1 + L$ where $L$ is a lighter tailed random variable (i.e., $\lim_{s \rightarrow \infty} s \prob(b(s)^{-1} L > r) = 0$), then $X_2 - X_1$ = $L$, but $L \notin RV_+^1(\alpha)$.
To construct a vector space, we will rely on a linear construction, and as we additionally wish to restrict attention to the positive orthant, we use the transformed-linear arithmetic of \cite{cooley2019decompositions}.

\subsection{Inner product space $\mathcal{V}^q$ and modeling subset $\mathcal{V}^q_+$}
\label{sec:innerProductSpace}
We define a vector space constructed from transformed-linear combinations of regularly varying random variables as
\begin{equation}
\label{eq:V}
\mathcal{V}^q = \big\{\bm a^\top \circ \bm Z = a_{1} \circ Z_{1} \oplus \cdots \oplus a_{q} \circ Z_{q}, \,\bm{a}\in \reals^q  \},
\end{equation}
where, as before, $\bm Z = (Z_1, \ldots Z_q)^\top\in RV_+^q(\alpha)$ is a vector of independent regularly varying random variables with normalizing function $b(s) \rightarrow \infty$ resulting in unit-scale for all $Z_j$, $j= 1,\ldots, q$, and which also meet the lower tail condition~\eqref{eq:lowertail}.
    
For $X_1 = \bm a_1^\top \circ \bm Z$ and $X_2 = \bm a_2^\top \circ \bm Z$ in $\mathcal{V}^{q}$, we define vector addition as $X_1 \oplus X_2 = (\bm a_1 + \bm a_2)^\top \circ \bm Z$ and scalar multiplication as $c \circ X_1 = c \bm a_1^\top \circ \bm Z,$ for $c \in \mathbb{R}$.
It follows that $\mathcal{V}^q$ is a vector space since it is isomorphic to $\mathbb{R}^q$; any $X \in \mathcal{V}^q$ is uniquely determined by its coefficient vector $\bm a$ \citep{lee2022phd}.
Since they follow standard arguments, the conditions required for $\mathcal{V}^q$ to form a vector space are shown in Section of 1.3 of the supplement.
Note that the vector space $\mathcal{V}^q$ differs from the vector space of \cite{cooley2019decompositions} in that elements of $\mathcal{V}^q$ are random variables.
Also note that $\Vq$ does not only consist of regularly-varying random variables.
The zero element $\bm 0^ \top \circ \bZ = t(\bm 0) \; a.s.$, and an element $X=\bm{a}^\top \circ \bZ$ with $a_j\le 0$ for all $j$ satisfies $\lim_{s \rightarrow \infty} s\prob(b(s)^{-1}X > x) = 0$ for all $x > 0$.

As $\mathcal{V}^q$ is isomorphic to $\mathbb{R}^q$, it is natural to define the inner product on $\mathcal{V}^q$ as
\begin{equation}
\label{eq:innerProduct}
    \langle X_{1}, X_{2} \rangle := \bm a_1^\top \bm a_2 = \sum_{i=1}^{q}a_{1i}a_{2i}.
\end{equation}
The question then becomes, why is this inner product meaningful for members of $\mathcal{V}^q$?
Below, we argue that this inner product does make sense when $\alpha=2$.
It is helpful to consider the {\em preimages} of the vector space's elements when discussing the inner product.

For each $X_i \in \mathcal{V}^q$, there is a preimage $Y_i = t^{-1}(X_i)$ which takes values in $\mathbb{R}$.
The `arithmetic' of transformed-linear operations is performed on this preimage space; for example, $X_1 \oplus X_2 = t(Y_1 + Y_2)$.
Save for the zero element, the preimage $Y_i$ is regularly varying whether $X_i$ is or not; for the time being consider $Y_i = t^{-1}(\bm a_i ^\top \circ \bZ)$ with $\bm a_i \neq \bm 0$.
With $\alpha = 2$, the scale of $Y_i$ is $\sum_{j = 1}^q a_{ij}^2 = \langle X_i, X_i \rangle$.
More importantly with $\alpha = 2$, the inner product can be connected to dependence of the preimages.
Let $\bY=(Y_1,Y_2)^\top:=(t^{-1}(X_1),t^{-1}(X_2))^\top$, which is jointly regularly varying with support $\reals^2$ and angular measure given by $H_{\bY}(\cdot)=\sum_{j=1}^{q}\|\bm a_{\cdot j}\|^2\delta_{\bm a_{\cdot j}/\|\bm a_{\cdot j}\|}(\cdot)$ for Borel sets in $\Theta^1=\{\bm v\in\reals^2:\|\bm v\|=1\},$ as shown in Section 1.4 of the supplement.
Then, 
\begin{equation*}
    \langle X_1, X_2\rangle
    =
    \sum_{j=1}^{q}a_{1j}a_{2j}
    =
    \sum_{j=1}^{q}\rbr{\frac{a_{1j}}{\|\bm a_{\cdot j}\|}}\rbr{\frac{a_{2j}}{\|\bm a_{\cdot j}\|}}\|\bm a_{\cdot j}\|^2
    =
    \int_{\Theta^1}v_1 v_2\diff H_{\bY}(\bm v),
\end{equation*}
which is analogous to (\ref{eq: TPDM}) with $\alpha = 2$.

Considering vectors $\bm X = (X_1, \ldots, X_p)^\top$ with $X_i \in \mathcal{V}^q$ for $i= 1, \ldots, p$,
we denote the corresponding matrix of inner products by
$
  \Gamma_{\bm X} = \langle X_i, X_k \rangle_{i,k\in\cbr{1, \ldots, p}} = A A^\top.
$
We say $X_1, X_2 \in \mathcal{V}^q$ are orthogonal if $\langle X_1, X_2 \rangle = 0$.
The norm on $\Vq$ is defined as $\| X \|_{\mathcal{V}^{q}} = \sqrt{ \langle X, X \rangle}$, whose subscript $\mathcal{V}^{q}$ distinguishes this norm from the usual Euclidean norm.
With norm, we can show that $\mathcal{V}^{q}$ is complete \citep{lee2022phd}, which follows from its isometry to $\mathbb{R}^q$, which is shown in Section 1.5 of the supplement for completeness.
Finally, the norm defines a metric $
d(X_1, X_2) := \| X_1 \ominus X_2 \|_{\mathcal{V}^{q}}=\sqrt{\sum_{j=1}^{q}(a_{1j}-a_{2j})^2}.
$

For modeling and prediction, we need to only consider a subset of $\mathcal{V}^q$.
As we assume $X_i = \bm a_i^\top \circ Z$ is regularly varying, $\max_{j = 1, \ldots, q} a_{ij} > 0$.
Furthermore, while negative coefficients are essential to form a complete vector space, their contributions are not reflected in the framework of regular variation.
That is, if $\bm X = A \circ \bm Z$ where $A$ has some negative elements, then its tail behavior is indistinguishable from $\bm X_+ = A^{(0)} \circ \bm Z$ as they have the same angular measure: $H_{\bm X} = H_{\bm X_+} = \sum_{j = 1}^q \|  \bm a^{(0)}_{\cdot j} \|^2 \delta_{ \bm a^{(0)}_{\cdot j} / \|  \bm a^{(0)}_{\cdot j} \|}(\cdot)$.
For modeling purposes, it is sufficient to restrict attention to the subset 
\begin{equation*}
\label{eq:V+}
    \mathcal{V}^q_+ = \big\{\bm a{^\top} \circ \bm Z = a_{1} \circ Z_{1} \oplus \cdots \oplus a_{q} \circ Z_{q}\},
\end{equation*}
where $a_j \geq 0$ for all $j = 1, \ldots, q$ and $a_j > 0$ for some $j = 1, \ldots, q$, and the random vector $\bm Z = (Z_{1}, \ldots Z_{q})^\top$ is defined as in (\ref{eq:V}).

A rationale for assuming $\mathcal{V}^q_+$ as a general modeling framework for prediction in the case of regular variation is as follows.
Consider a general $\bX\in RV_+^p(2)$ with any angular measure $H_{\bX}$, not necessarily in $\mathcal{V}^q$.
Even when the angular measure $H_{\bm X}$ is unattainable, the tail dependence of $\bX$ can be summarized by the TPDM $\Sigma_{\bm X}$, readily estimated from realizations of $\bX$.
Since $\alpha = 2$, $\Sigma_{\bm X}$ is completely positive \citep[][Section 4]{cooley2019decompositions}; that is, there exists a $p \times q^*$ nonnegative matrix $A^*$, with $q^* < \infty$, such that $\Sigma_{\bm X} = A^* {A^*}^\top$.
Define $\bm X^* = A^* \circ \bm Z$, with elements $X_i^* \in \mathcal{V}^q_+$ for $i=1,\ldots,p.$
Through the lens of the TPDM, $\bX^*$ can be viewed as the `twin' of $\bX$ as they share the same tail behavior as captured by the TPDM. 
Prediction in the next section is based only on $\Sigma_{\bm X} = \Gamma_{\bm X^*}$; and the underlying $q^*$ and $A^*$ will not need to be estimated.
Thus, for modeling and prediction purposes, we assume $X_i\in \mathcal{V}^q_+$ for $i=1,\ldots,p$, throughout.

\subsection{Transformed-linear prediction via projection theorem}
\label{sec:transLinearPred}

As $\mathcal{V}^q$ is isomorphic to Hilbert space $\mathbb{R}^q$, the best transformed-linear predictor can be derived analogously to the BLUP in classical linear prediction.
Assume $X_i = \bm a_i^\top \circ \bm Z \in \mathcal{V}^q$ for $i = 1, \ldots, p+1$, and let $\bm X_p = (X_1, \ldots, X_p)^\top$.
We aim to find $\bm b \in \mathbb{R}^p$ that minimizes $d(\bm b^\top \circ \bm X_p, X_{p+1})$. 
Writing in matrix form
\begin{equation*}
\begin{bmatrix}
\bm{X}_{p}\\
X_{p+1}
\end{bmatrix}
=
\begin{bmatrix}
A_p\\
\bm{a}_{p+1}^\top
\end{bmatrix}
\circ \bm{Z},
\end{equation*}
where $A_{p}= (\bm a_1^\top, \ldots, \bm a_p^\top)^\top$, the inner product matrix of $(\bm{X}_{p}^\top, X_{p+1})^\top$ via~\eqref{eq:innerProduct} is
\begin{equation}\label{eq:GammaMatrices}
\Gamma_{(\bm{X}_{p}^\top, X_{p+1})^\top}
=
\begin{bmatrix}
A_{p}A_{p}^\top & A_{p}\bm{a}_{p+1}\\
\bm{a}_{p+1}^\top A_{p}^\top & \bm{a}_{p+1}^\top\bm{a}_{p+1}
\end{bmatrix}
:=
\begin{bmatrix}
\Gamma_{11} & \Gamma_{12}\\
\Gamma_{21} & \Gamma_{22}
\end{bmatrix}.
\end{equation}
Minimizing $d(\bm b^\top \circ \bm X_p, X_{p+1})$ is equivalent to minimizing $\|\bm{b}^\top A_p-\bm{a}_{p+1}^\top\|_{2}^{2}$.
Taking derivatives with respect to $\bm b$ and setting them equal to zero, the minimizer $\bm b^*$ solves
$(A_{p}A_{p}^\top)\bm b^*=A_{p}\bm{a}_{p+1}$. 
If $A_{p}A_{p}^\top$ is invertible, then the solution $\bm{b}^*$ is,
\begin{equation}\label{eq: bHat}
\bm{b}^*=(A_{p}A_{p}^\top)^{-1}A_{p} \bm{a}_{p+1}=\Gamma_{11}^{-1}\Gamma_{12}.
\end{equation}
An equivalent way to derive the best transformed-linear prediction is through the projection theorem \citep{lee2022phd}. 
\begin{theorem}
(Projection theorem in $\mathcal{V}^q$)\label{ProjThm}
Assume $X_i = \bm a_i^\top\circ \bZ \in \mathcal{V}^q$ for $i= 1, \ldots, p$ with $p < q$.
Let $\bX_p=(X_1,\ldots,X_p)^\top$.
Consider the closed subspace 
$\mathcal{V}_{A}=\{\bm{b}^\top\circ\bX_p; 
\bm{b}\in \mathbb{R}^{p}\}$;
that is, the subspace of $\Vq$ spanned by $\{X_{1},\ldots,X_{p}\}$. 
Define the orthogonal complement 
$\mathcal{V}_{A}^{\perp}=\{X\in \mathcal{V}^{q}; \langle X,Y\rangle=0, \,\forall\, Y\in\mathcal{V}_{A}\}.$
For $X\in \mathcal{V}^{q}$,
\begin{enumerate}
    \item There exists a unique element $\hX \in \mathcal{V}_A$ such that
\[
\|X\ominus\hX\|_{\mathcal{V}^{q}}=\inf_{Y\in\mathcal{V}_{A}}\|X\ominus Y\|_{\mathcal{V}^{q}}, \text{ and }
\]
    \item $\hX\in \mathcal{V}_{A}$ such that $\|X\ominus \hX\|_{\mathcal{V}^{q}}=\inf_{Y\in\mathcal{V}_{A}}\|X\ominus Y\|_{\mathcal{V}^{q}}$ if and only if $\hX \in \mathcal{V}_{A}$ and $(X\ominus\hX)\in \mathcal{V}_{A}^{\perp}$.
\end{enumerate}
\end{theorem}
Proofs are provided in Section 1.6 of the supplement.
By Theorem \ref{ProjThm}, the best transformed-linear predictor $\hX_{p+1}={\bm{b}^*}{^\top} \circ \bm{X}_p$ is such that the prediction error $X_{p+1}\ominus\hX_{p+1}$ is orthogonal to the subspace spanned by $X_1,\ldots, X_p$.
The orthogonality condition can be stated as $\langle X_{p+1}\ominus\hX_{p+1},X_i\rangle=0$ for all $i=1, \ldots, p$.
By linearity of inner products, this is equivalently written as
\begin{align*}
\begin{bmatrix}
\langle X_{p+1},X_{i}\rangle\\
\end{bmatrix}_{i=1}^{p}
&=
\begin{bmatrix}
\langle X_{i},X_{j}\rangle
\end{bmatrix}_{i,j=1}^{p}
\begin{bmatrix}
b_{i}
\end{bmatrix}_{i=1}^{p}
=
\begin{bmatrix}
\sum_{k=1}^{q}a_{ik}a_{jk}
\end{bmatrix}_{i,j=1}^{p}
\begin{bmatrix}
b_{i}
\end{bmatrix}_{i=1}^{p}.
\end{align*}
By (\ref{eq:GammaMatrices}), $\bm{b}^*$ satisfies $(A_{p}A_{p}^\top)\bm{b}^*=A_{p}\bm{a}_{p+1}$ as above.

The metric being minimized by (\ref{eq: bHat}) is $d(\widehat X_{p+1}, X_{p+1}) = \| \widehat X_{p+1} \ominus X_{p+1} \|_{\mathcal{V}^{q}} = \sum_{j = 1}^q (a_{1j} - a_{2j})^2$, which can be understood as the scale of the difference between the {\em preimages} $t^{-1}(\widehat{X}_{p+1})$ and $t^{-1}(X_{p+1})$.
It is reasonable to ask what is being minimized in the space in which the $X_i$'s live.
The scales of $\widehat X_{p+1} \ominus X_{p+1}$ and $X_{p+1} \ominus \widehat X_{p+1}$ are $\sum_{j = 1}^q \left( (a_{1j} - a_{2j})^{(0)} \right)^2$ and $\sum_{j = 1}^q \left( (a_{2j} - a_{1j})^{(0)} \right)^2$, respectively, and these scales are generally not equal due to the zero operation. 
Nevertheless, the scale of $ \max \cbr{ \widehat X_{p+1} \ominus X_{p+1}, X_{p+1} \ominus \widehat X_{p+1}}$ is $\sum_{j = 1}^q (a_{1j} - a_{2j})^2 = d^2(\widehat X_{p+1}, X_{p+1})$, which can be estimated from the realizations.
A proof is provided in Section 1.7 of the supplement.


\section{Uncertainty quantification for prediction error}
\label{sec:UQ}

In the non-extreme setting, linear prediction minimizes MSPE.
As MSPE corresponds to the conditional variance under a Gaussian assumption, it is used to construct prediction intervals.
Similarly, our transformed-linear predictor minimizes
\begin{equation*}\label{eq:gamma2.1}
\begin{split}
\|\hX_{p+1} \ominus X_{p+1}\|^{2}_{\mathcal{V}^{q}}
&=({\bm b^*}^\top A_{p}-\bm{a}_{p+1}^\top)({\bm b^*}^\top A_{p}-\bm{a}_{p+1}^\top)^\top\\
&=\Gamma_{22}-\Gamma_{21}\Gamma_{11}^{-1}\Gamma_{12}=:\Gamma_{2|1},\\
\end{split}
\end{equation*}
where $\Gamma_{ij}$ for $i,j=1,2$ as in \eqref{eq:GammaMatrices}.
Unlike MSPE, $\Gamma_{2|1}$ is not understood via expectation, but instead as the scale of 
$
  \max \cbr{( X_{p+1} \ominus \hX_{p+1}), (\hX_{p+1} \ominus X_{p+1})}.
$
Despite its similarity to MSPE, $\Gamma_{2|1}$ may not be suitable for constructing prediction intervals.
Unlike the Gaussian case, where the conditional variance does not depend on the predicted value $\hX_{p+1}$, the error magnitude in the polar geometry of regular variation is conditional on the size of the predicted value $\hX_{p+1}$.
A visual illustration of this is provided in Section 2.1 of the supplement.
To address this, we first characterize the tail dependence between $\hX_{p+1}$ and $X_{p+1}$ and then use the polar geometry of regular variation to construct prediction intervals when $\hX_{p+1} = x$ is large in Section~\ref{sec:condtlInterval}.


\subsection{Constructing angular measures via completely positive decomposition}
\label{sec:constructAngviaCPD}


Describing the prediction error requires characterization of the tail dependence between the predictor $\hX_{p+1}$ and predictand $X_{p+1}$.
As the angular measure $H_{(\hX_{p+1},X_{p+1})}$ completely describes the asymptotic tail dependence, if it were accessible, it could be used to provide a prediction interval.
However, since we assume the angular measure of $(\bX_p^\top, X_{p+1})^\top$ is unattainable, we do not have direct access to $H_{(\hX_{p+1},X_{p+1})}$.

Continuing to assume $X_i\in\Vq$ for $i = 1, \ldots, p+1,$ writing our vector in matrix form yields
\begin{equation*}
\label{eq:matrix C}
    \begin{bmatrix}
        \widehat{X}_{p+1}\\
        X_{p+1}
    \end{bmatrix}
    =
    \begin{bmatrix}
        {\bm b^*}^\top A_p\\
        \bm a_{p+1}^{\top}
    \end{bmatrix}
    \circ
    \bm Z,
\end{equation*}
where the random vector $\bm Z = (Z_{1}, \ldots Z_{q})^\top$ is defined as in (\ref{eq:V}).
The $2 \times 2$ {\em prediction} inner product matrix is then defined as
\begin{equation}\label{eq:predTPDM}
\begin{split}
\Gamma_{(\hX_{p+1}, X_{p+1})}
&=
\begin{bmatrix}
{\bm b^*}^\top{A}_p\\
\bm{a}_{p+1}^\top
\end{bmatrix}
\begin{bmatrix}
{A}_{p}^\top{\bm b^*} &
\bm{a}_{p+1}
\end{bmatrix}
=
\begin{bmatrix}
\Gamma_{21}\Gamma_{11}^{-1}\Gamma_{12} & \Gamma_{21}\Gamma_{11}^{-1}\Gamma_{12}\\
\Gamma_{21}\Gamma_{11}^{-1}\Gamma_{12} & \Gamma_{22}
\end{bmatrix}
\end{split},
\end{equation}
which, under the assumption that $X_1, \ldots, X_p, X_{p+1} \in \mathcal{V}^q_+$, is directly obtainable from the block matrices of $\Sigma_{(\bX^\top, X_{p+1})^\top}$, or, in practice, from its estimate, with the estimated TPDM $\widehat{\Sigma}_{ij}$ replacing the corresponding $\Gamma_{ij}$.

To obtain a prediction interval, we obtain a candidate angular measure $\widetilde H_{(\hX_{p+1},X_{p+1})}$ which agrees with the information we have in $\Gamma_{(\hX_{p+1}, X_{p+1})}$. Importantly, we only need to characterize the bivariate relationship between $\hX_{p+1}$ and  $X_{p+1}$, and our candidate angular measure exists on $\Theta^1_+$.
We obtain $\widetilde H_{(\hX_{p+1},X_{p+1})}$ via a completely positive decomposition of $\Gamma_{(\hX_{p+1}, X_{p+1})}$ yielding an angular measure initially consisting of many point masses, which we will later smooth to be continuous.

$\Gamma_{(\hX_{p+1}, X_{p+1})}$ is known to be completely positive, since it is a $2\times2$ positive definite matrix with positive entries.
As with other linear prediction like simple kriging, our weight matrix $\bm b^*$ can contain negative entries, and consequently it is possible that ${\bm b^*}^\top A_p$ could contain negative entries and $\widehat X_{p+1} \notin \Vq_+$ (even though $\widehat X_{p+1} \geq 0 \; a.s.$).
Nevertheless, $\Gamma_{(\hX_{p+1}, X_{p+1})}$'s off-diagonal entry is $\Gamma_{21}\Gamma_{11}^{-1}\Gamma_{12}$ which is greater than zero since $\Gamma_{11}$, the inner product matrix of $\bm X_p$, is positive definite. 

Given a $p \times p$ completely positive matrix $S$, completely positive decomposition seeks to find a $p \times q^*$ nonnegative matrix $C$ such that $S = C C^\top$, \cite{dur2010} provides a survey of the topic.
The matrix $C$ is not unique and the minimum value of $q^*$ for a given $p$ is not known.
For a $2\times2$ matrix 
$S = 
{\tiny
\left( 
  \begin{array}{r r}
  a & b\\
  b & c
  \end{array}
\right)
}
$, there is a known $2\times2$ solution 
$C = 
{\tiny
\left( 
  \begin{array}{r r}
  \sqrt{a}   & 0\\
  b/\sqrt{a} & \sqrt{c - b^2/a}
  \end{array}
\right)
}
$.
However, since we aim to construct a candidate angular measure consisting of many point masses, we employ an existing algorithm \citep{groetzner2020} for obtaining a decomposition matrix $C$ with moderate sized $q^*$, and repeat the decomposition procedure over $n_{de}$ iterations.
The approach exploits the non-uniqueness of the completely positive decomposition, yielding a distinct nonnegative matrix $C_l$ for each iteration $l = 1, \ldots, n_{de}$.  
The number of iterations, $n_{de}$, can be chosen as large as necessary to achieve the desired level of continuity in the angular measure.
We then define the normalized angular measure as
\begin{equation}
\label{eq:empiricalAngularMeasure}
    \widetilde{H}_{(\hX_{p+1}, X_{p+1})} = \frac{1}{n_{de}} \sum_{l = 1}^{n_{de}} H_{C_l \circ \bm Z}.
\end{equation}
The normalized angular measure $\widetilde{H}_{(\hX_{p+1}, X_{p+1})}$ consists of $n_{de}\times q^*$ point masses.

\subsection{Prediction intervals for $X_{p+1}$ given large $\hX_{p+1}$}
\label{sec:condtlInterval}

Our primary interest lies in the conditional behavior of $X_{p+1}$ given a specific large value $\hX_{p+1} = x$.
To construct these prediction intervals, we modify the approach of \cite{cooley2010pairwise}, who used the limiting intensity function of regular variation to approximate the density of $X_{p+1}$ given large values of $\bm X_p = \bm x_p$.
\cite{cooley2010pairwise} begin by transforming (\ref{eq:nu}) from polar to Cartesian coordinates.
This transformation which has Jacobian $|J|=\|\bm{x}\|_2^{-(p+1)}x_{p+1}$ when the angular measure is defined in terms of the $L_2$ norm \citep{song1997}, which we will assume henceforth.
Assuming $H_{\bm X}$ is continuously differentiable, they obtain the limiting measure's intensity function
\[
\nu_{(\bm X_p^\top, X_{p+1})^\top}(\bm x)\diff\bm x = \alpha  \|\bm x \|_2^{-(\alpha + p+2)} x_{p+1} h_{(\bm X_p^\top, X_{p+1})^\top}( \bm x \|\bm x\|_2^{-1}) \diff \bm x,
\]
and the approximate conditional density is 
\[
f_{X_{p+1}|\bm X_p}(x_{p+1} | \bm x_p) \approx c^{-1} \nu_{(\bm X_p^\top, X_{p+1})^\top}( \bm x_p, x_{p+1}),
\]
where $c = \int_0^\infty \nu_{(\bm X_p^\top, X_{p+1})^\top}(\bm x) \diff x_{p+1}.$
\cite{cooley2012approximating} fit a parametric model for the angular density $h_{(\bm X_p^\top, X_{p+1})^\top}$ and applied their method in dimension ($p = 4$); applying the approach for larger $p$ would require a high-dimensional angular measure model.


In this work the problem is simplified; regardless of $p$, we only need to describe a bivariate relationship between $\hX_{p+1}$ and $X_{p+1}$, and we can let $\alpha = 2$ and $p = 1$ in the above expressions.
To find the bounds of the prediction interval for a given value $\hat x_{p+1}$, we seek a value $u$ 
such that 
\begin{equation}
    \label{eq:predInterval}
      \rho 
  = \int_0^u f_{X_{p+1}|\hX_{p+1}}(x_{p+1}|\hat x_{p+1})\diff x_{p+1} 
  =\frac
    {\int_0^u 2 \|\bm x\|_2^{-5} x_{p+1} h(\bm x \|\bm x\|_2^{-1}) \diff x_{p+1}}
    {\int_0^\infty 2 \|\bm x\|_2^{-5} x_{p+1} h(\bm x \|\bm x\|_2^{-1}) \diff x_{p+1}},
\end{equation}
where $\bm x = (\hat x_{p+1}, x_{p+1})^\top,$ and $\rho$ denotes the prediction level; e.g., setting $\rho=0.025$ and 0.975 yields a central 95\% prediction interval of $x_{p+1}$ given $\hat x_{p+1}$.
We apply weighted kernel density estimation to obtain a smooth, nonparametric estimate of $h$, using the angular point masses derived from the completely positive decomposition, as described in Section~\ref{sec:constructAngviaCPD}.
The weighted kernel density estimate of $h$ in terms of the angular component is
\begin{equation*}
    \hat{h}(w)=\sum_{l=1}^{n_{de}}\sum_{j=1}^{q^*}m_{j,l}\phi_{b}(w-\delta_{j,l}), \quad w\in\Theta_+^1,
\end{equation*}
where the angular weight is $m_{j,l}:=n_{de}^{-1}\|\bm{c}_{j,l}\|_2^2$ with $\bm{c}_{j,l}$ denoting the $j$th column of the $l$th nonnegative matrix $C_l$.
The corresponding angular coordinate is $\delta_{j,l}=[c_{j,l}]_1/\|\bm{c}_{j,l}\|_2$, $j= 1,\ldots,q^*$, $l=1,\ldots,n_{de}$, and $\phi_b$ is a scaled kernel with bandwidth $b$, e.g., a Gaussian kernel, $\phi_b(s)=b^{-1}\phi(s/b).$
In the simulation study and application below, we choose the bandwidth via a default method (e.g., the \texttt{kde} function in the \texttt{ks} R package), or through a $k$-fold cross-validation as described below.
For the kernel estimation, we employ the adjusted boundary bias approach of \cite{marron1994transformations} to account for the bounded support of the angular density $h$.

\section{Implementation and Simulation study}
\label{sec:simulation}

Implementing extremal transformed linear prediction begins with estimation of the TPDM, which we do pairwise.
If $(X_i, X_j)$ is a two-dimensional marginal vector of $\bX\in RV^p_+(\alpha)$,
\cite{mhatre2021transformed} show that $\sigma_{ij} = \int_{\Theta_+^{p-1}} w_i^{\alpha/2} w_j^{\alpha/2} \diff H_{\bX}(\bm w) = \int_{\Theta_+^{1}} w_i^{\alpha/2} w_j^{\alpha/2} \diff H_{(X_i, X_j)}(w_i, w_j)$, when same normalizing function $b(s)$ is used to define both $H_{\bX}$ and $H_{(X_i, X_j)}$. 
Consider $\bm X \in RV^2_+(2)$.
For a given norm, define the radial component as $R=\|(X_i,X_j)\|$ and the angular component as $(W_i,W_j)=(X_i,X_j)/R$ for $i,j\in \{1,\ldots,p\}$.
Then, the $(i,j)$th element of the TPDM can be equivalently expressed as $\sigma_{ij}=\lim_{s\rightarrow\infty}mE[W_j W_j\,|\, R>s]$, where $m:=H_{\bX}(\Theta_+^1)$ is the total mass of the angular measure. 
This leads to a natural estimator. 
Letting $(X_{it},X_{jt})$, $t=1,\ldots,n,$ be i.i.d.\ copies of $(X_i,X_j)$ and defining $R_t=\|(X_{it},X_{jt})\|_2$, and $(W_{it},W_{jt})=(X_{it},X_{jt})/R_{t}$, define
\begin{equation}
\label{eq:TPDMestimator}
    \hat{\sigma}_{ij}(n,k)=\frac{m}{k}\sum_{t=1}^{n}W_{it}W_{jt}\1[R_{t}>R_{(k)}],
\end{equation}
where $\1(\cdot)$ denotes the indicator function and $R_{(k)}$ is the $k$th largest order statistic of $\{R_t\}_{t=1}^n.$
If $k:=k(n)$ satisfies $k\rightarrow \infty$ and $k/n\rightarrow 0$ as $n\rightarrow \infty$, \cite{larsson2012} show that this empirical estimator is consistent and asymptotically normal, under suitable conditions.
Assuming $\alpha=2$ and choosing the $L_2$-norm provides convenient properties, e.g., when the data are preprocessed to have a common unit scale, the total mass becomes $m=p$, removing the need for its estimation.


Positive dependence in $\bX$ is induced via matrix multiplication as in~\eqref{eq:linConst}. 
Let $(p+1,q)$ denote the dimension of the matrix $A$, and let $n$ denote the sample size.
We set $(p+1,q)=(7,400)$ and let $\bX:=(\bX_p^\top,X_{p+1})^\top\in\Vq_+$, where $\bm Z$ consists of $q$ independent unit-scale Pareto variables with $\alpha=2$, and $A$ is a nonnegative matrix whose entries are a specific realization from a uniform distribution over $[0,5]$. 
Each row of $A$ is normalized to unit norm.
The angular measure $H_{A\circ \bm Z}$ thus has 400 point masses, and the specified TPDM is $\Sigma_{\bX}=AA{^\top}$ from which we obtained the vector $\bm b^*=\Sigma_{11}^{-1}\Sigma_{12}$.

\begin{figure}[ht]
\centering
\includegraphics[width=2in,height=\textheight]{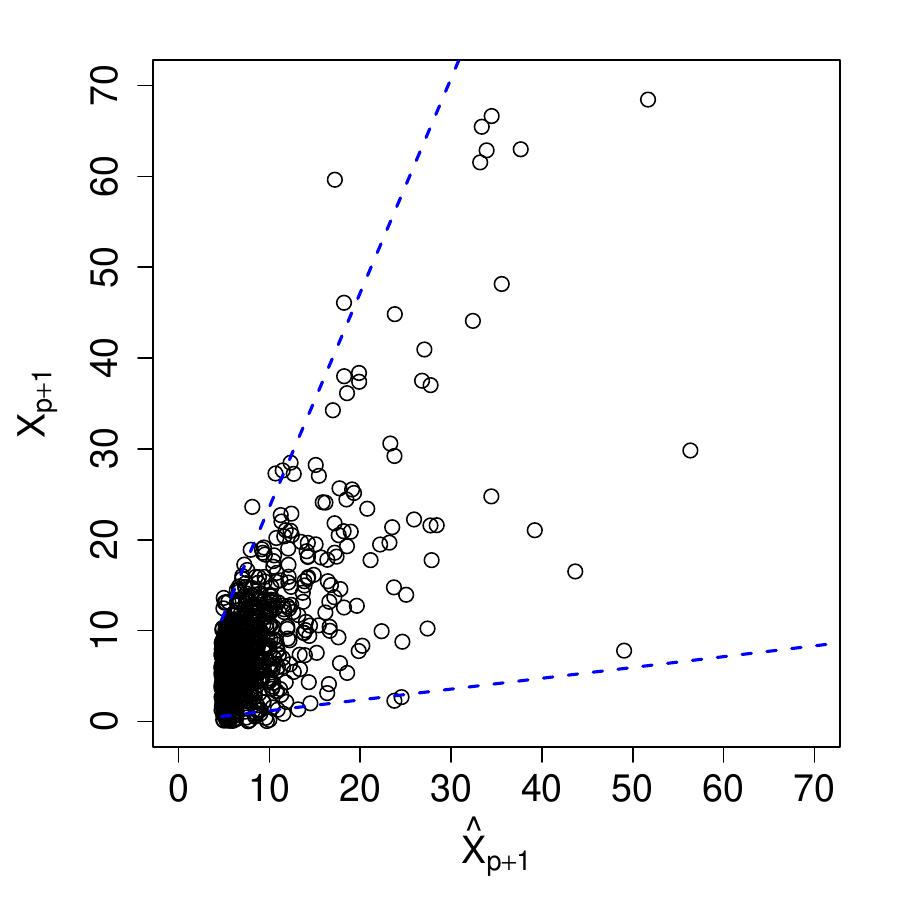}\includegraphics[width=2in,height=\textheight]{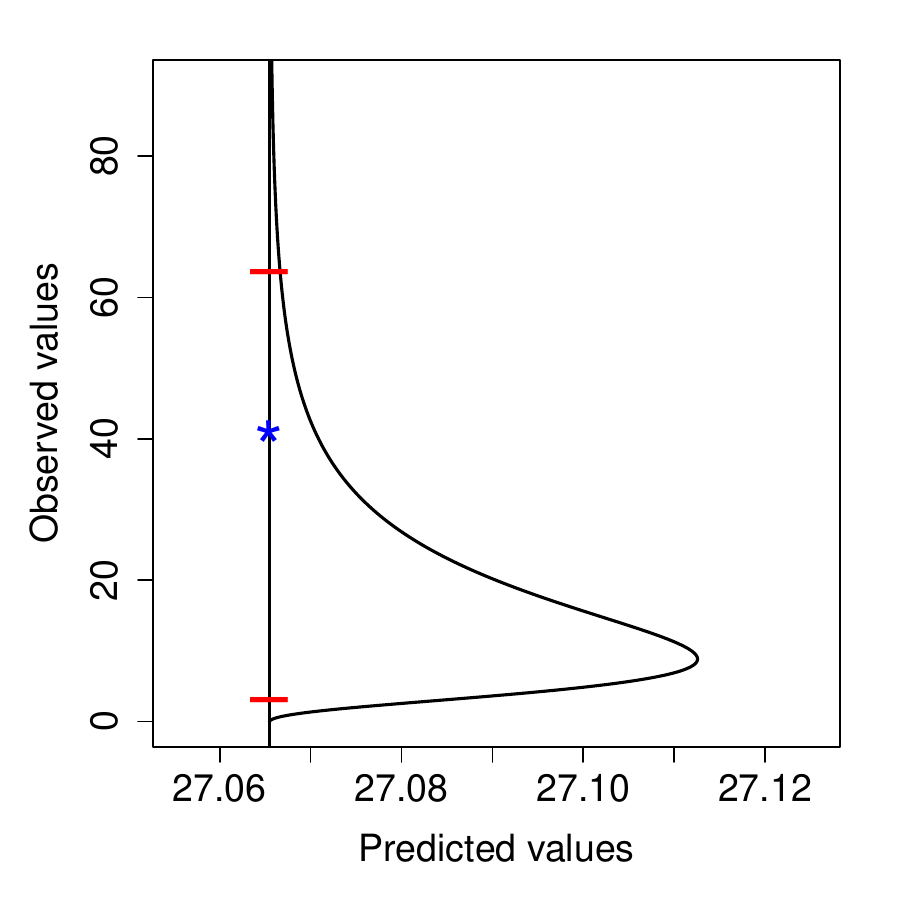}
\caption{
(Left) the scatter plot of $\hX_{p+1}$ and $X_{p+1}$ with $95\%$ conditional prediction intervals given each large value of $\hX_{p+1}.$
(Right) The approximate conditional density $f_{X_{p+1}|\hX_{p+1}}(X_{p+1}|\hX_{p+1} = 27.07)$.
The horizontal segments indicate the $95\%$ conditional prediction interval, and the star denotes the actual value of 40.95.
The units of the horizontal axis are the predicted values and the units of the conditional density are omitted.
}
\label{fig:simStudy}
\end{figure}

We generate 60,000 random realizations of $\bX=A\circ \bm Z$, using the first 40,000 samples as a training set for TPDM estimation.
We set the 0.75 quantile of $R_t = \|(X_{it},X_{jt})\|_2$, and obtain an estimate $\widehat{\Sigma}_{\bX}$. 
The left panel of Figure~\ref{fig:simStudy} shows a scatterplot of $x_{p+1}$ vs $\hat x_{p+1}$ from the test set, along with the upper and lower bounds obtained from \eqref{eq:predInterval}, which were obtained with the default \texttt{kde} method.
The fact that the points lie within the interior indicates a strong relationship between $\hX_{p+1}$ and $X_{p+1}$, and clearly the width of the prediction interval needs to increase with $\hat x_{p+1}$.
The overall coverage rate of these intervals is $0.956$.
The right panel of Figure \ref{fig:simStudy} illustrates procedure for generating the prediction interval.
The conditional density for a particular realization where $\hat x_{p+1}=27.07$ is shown vertically along with its .025 and .975 quantiles used for the prediction interval, and the actual value $x_{p+1}=40.95$ denoted by the star.

Given the prediction TPDM estimate, we assess the variability of the decomposition process by repeatedly performing the completely positive decompositions and constructing the corresponding normalized angular measures, each consisting of 1,000 point masses.
We find the variability of the the empirical bounds $(w_{0.025},w_{0.975})$ to be low.
Consequently, subsequent analyses focus on variability arising from the estimated TPDM itself rather than the completely positive decomposition.
Further details are provided in Section 2.2 of the supplement.


\section{Applications}
\label{sec:applications}

\subsection{Nitrogen dioxide air pollution}

NO$_2$ is one of six air pollutants for which the US Environmental Protection Agency (EPA) has national air quality standards.
We analyze daily EPA NO$_2$ data\footnote{https://www.epa.gov/outdoor-air-quality-data/download-daily-data} from five locations in the Washington DC metropolitan area (see Figure \ref{fig: washingtonDC}).
The first four stations (McMillan 11-001-0043,  River Terrace 11-001-0041, Takoma 11-001-0025, Arlington 51-013-0020) have long data records spanning 1995-2020.
Alexandria does not have observations after 2016.
We will perform prediction at Alexandria given data at the other four locations.
Observations in Alexandria actually come from two different stations: 51-510-0009 which has measurements from January 1995 to August 2012 and 51-510-00210 from August 2012 to April 2016.
Exploratory analysis did not indicate any detectable change point in the Alexandria data either with respect to the marginal distribution or with dependence with other stations, so we treat this data as coming from a single station.
There are 5163 days between 1995 and 2014 where all five locations have measurements.
Because NO$_2$ levels have decreased over the study period, we detrend at each location using a moving average mean and standard deviation with window of 901 days to center and scale.

Our framework assumes each $X_i \in \Vq_+$, and the detrended NO$_2$ data must be transformed to meet this assumption.
It is unclear whether the NO$_2$ data is heavy-tailed.
Nevertheless, we believe the regular variation framework is useful for describing the tail dependence for this data after marginal transformation.
Characterizing dependence after marginal transformation is justified by Sklar's theorem \citep{sklar1959}, and such transformations are regularly used in multivariate extremes studies, see also Proposition 5.15 of \cite{resnick2008extreme}.
After viewing standard diagnostic plots, we fit a generalized Pareto distribution above each location's 0.95 quantile and obtain the marginal estimated cdf's $\widehat F_i$ which are the empirical cdf below the 0.95 quantile and the fitted generalized Pareto above.
Letting $X_i^{(orig)}$ denote the random variable for detrended NO$_2$ at location $i$, we define $X_i = (1-\widehat{F}_i(X_i^{(orig)}))^{-1/2}-\delta$ obtaining a `shifted' Pareto distribution for $i = 1, \ldots, 5$.
The shift $\delta = 0.9352$ is such that $\E[t^{-1}(X_i)]$ = 0.
This adjustment centers the preimages of the transformed data, which we found leads to reduced bias in estimating the TPDM.
We let $\bm X_t=(X_{1,t},\ldots,X_{5,t})^\top$ denote the random vector of observations on day $t$, which we assume to be i.i.d. copies of $\bm X$.
This is a simplifying assumption as there is some temporal dependence in the NO$_2$ data, but observations from previous days seem to add little if any information not contained in the day's observations from nearby stations used to perform prediction.  

We first conduct predictions for the period prior to 2015, allowing us to assess performance using observed data at Alexandria.
Indices are randomly drawn to divide the data set into training and test sets consisting of 3,442 and 1,721 observations respectively, and both sets cover the entire observational period.
Using the training set, we estimate the five-dimensional TPDM, $\widehat\Sigma_{\bm X}$, via pairwise estimation as described in Section~\textcolor{red}{\ref{sec:simulation}}.
Let $\bm x_t$ denote the observed measurements on day $t$.
For each $i \neq j$ in $1, \ldots, 5$, let $r_{t,ij} = \| (x_{t,i}, x_{t,j}) \|_2$ and $(w_{t,i},w_{t,j})=(x_{t,i},x_{t,j})/r_{t,ij}$.
We let $\hat{\sigma}_{ij}=2 n_{exc}^{-1}\sum_{t=1}^{n}{w_{t,i}w_{t,j}\1(r_{t,ij}>r^*_{ij})}$, where $n_{exc} = \sum_{t=1}^{n} \1(r_{t,ij}>r^*_{ij})$.
We choose $r^*_{ij}$ to correspond to the 0.95 quantile.
The constant 2 arises from knowledge that $m = 2$ due to the marginal transformation.

\begin{figure}[t]
\centering
\includegraphics[width=2in,height=\textheight]{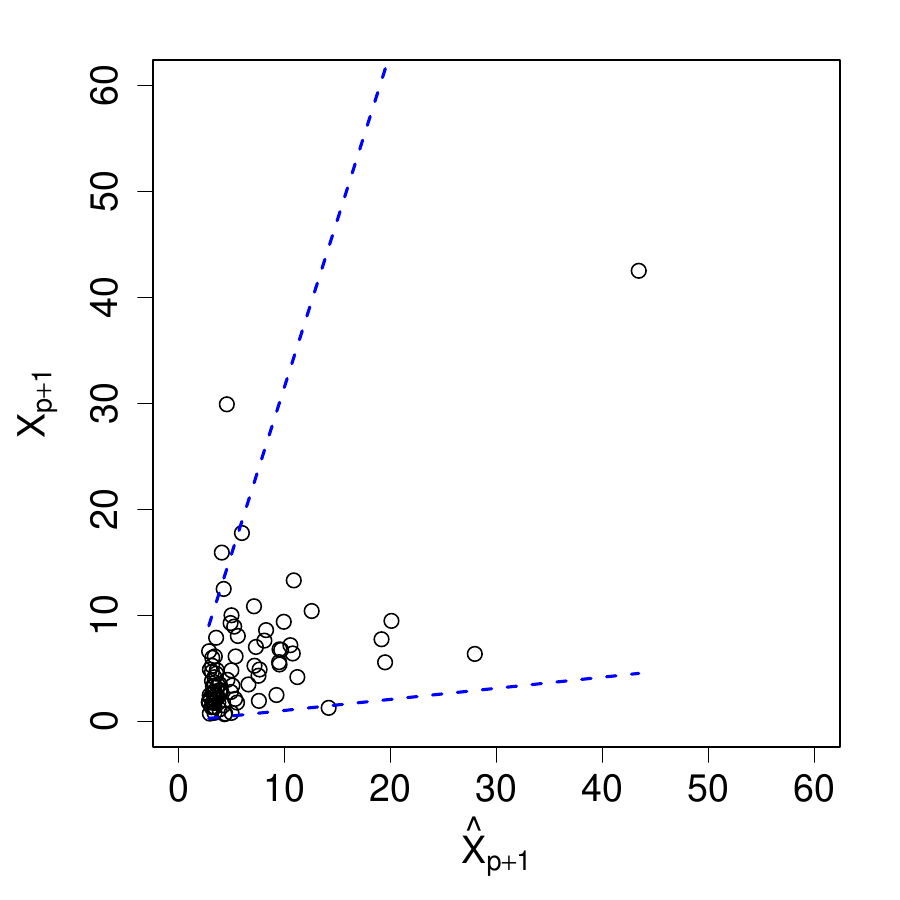}
\includegraphics[width=2in,height=\textheight]{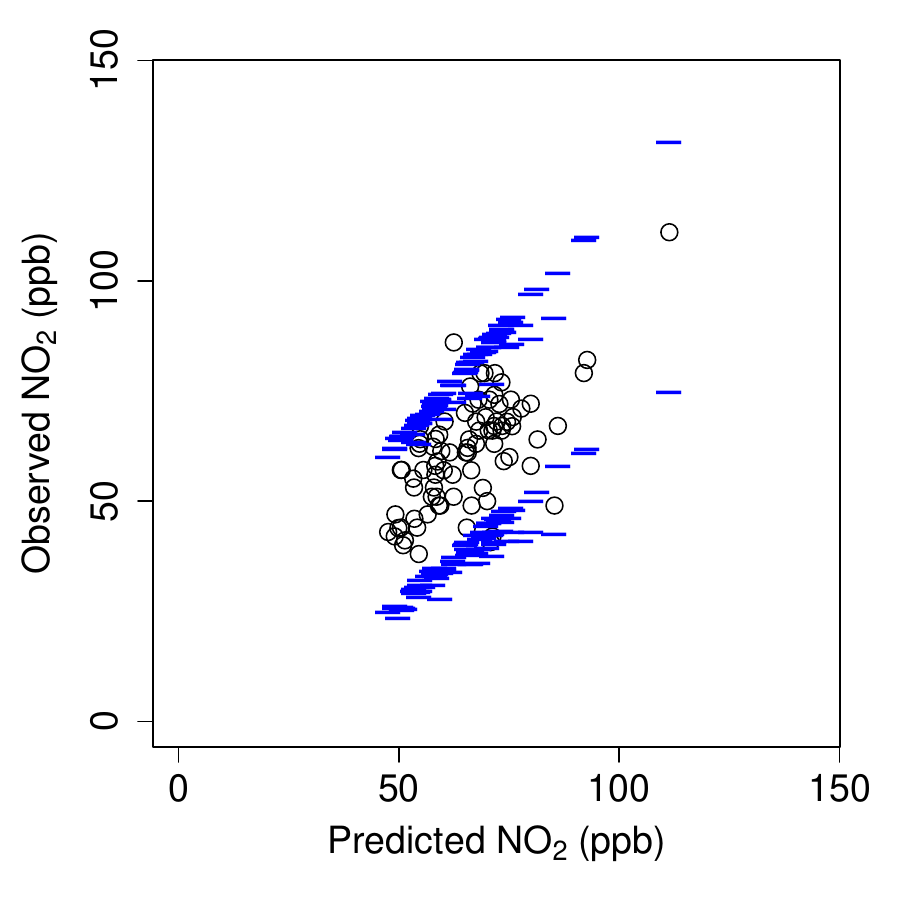}
\includegraphics[width=2in,height=\textheight]{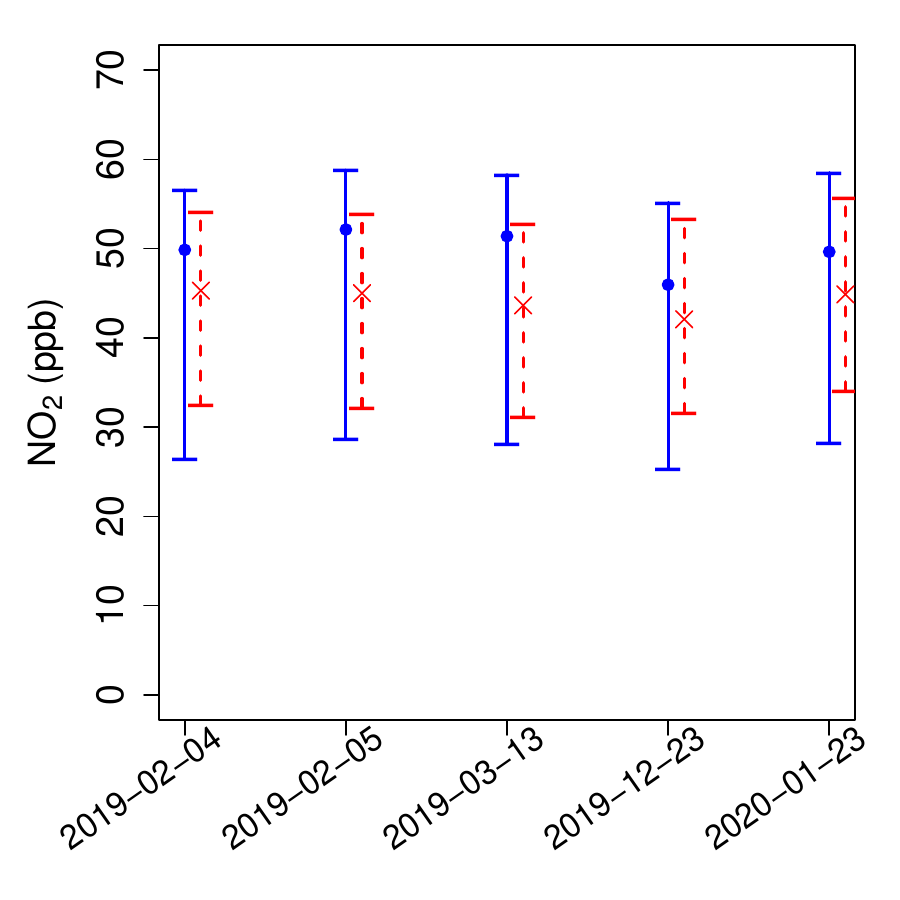}
\caption{(Left) Scatterplot of $\hX_{5}$ and $X_{5}$ with the 95\% prediction intervals on the Pareto scale. 
(Middle) Scatterplot and 95\% prediction intervals after transformation to the original scale of the NO$_2$ data.
(Right) Comparison of the point predictions and 95\% prediction intervals using the transformed linear approach (solid line) and a Gaussian-based approach (dashed line) for five recent dates when Alexandria is not observed.}
\label{fig:no2}
\end{figure}

From $\widehat\Sigma_{\bm X}$, we obtain $\hX_{t,5} = \hat{\bm b}^{*}{^\top} \circ \bm X_{t,4}$, where the estimated coefficient vector is $\hat{\bm b}^{*} = (0.096, 0.225, 0.123, 0.410)^\top$.
The order of the coefficients correspond to McMillan, River Terrace, Takoma, and Arlington.
We note that Arlington has the largest weight, which aligns with its geographic proximity to Alexandria.
From this, we obtain the prediction TPDM estimate with an off-diagonal element of 0.563.
We calculate $\hX_{t,5}$ for all $t$, but only consider those for which $\hX_{t,5}$ exceeds the 0.95 quantile.
The left panel of Figure \ref{fig:no2} shows the scatterplot of the observed values $x_{t,5}$ versus the predicted values $\hat x_{t,5}$.
By taking the inverse of the marginal transformation, multiplying by the moving average standard deviation and adding the moving average mean, the predicted value can be put on the scale of the original data.
The center panel of Figure \ref{fig:no2} shows the corresponding scatterplot on the original scale.

We use the method described in Section \ref{sec:constructAngviaCPD} to approximate $H_{(\hX_{p+1}, X_{p+1})}$.
Setting $q^*=10$, we iterate the decomposition to the prediction TPDM estimate $\widehat{\Sigma}_{(\widehat{X}_{p+1},X_{p+1})}$.
Given a specific normalized angular measure obtained from the decomposition, we use the method described in Section \ref{sec:condtlInterval} to create 95\% prediction intervals for each large predicted value $\hat x_{t,5}$, obtained using the default \texttt{kde} method.
Prediction intervals on the Pareto scale are shown in the left panel of Figure \ref{fig:no2}, with the coverage rate of these intervals is 0.965.
The intervals can similarly be back-transformed to be on the original scale as shown in the center panel of Figure \ref{fig:no2}.
The lack of monotonicity in these intervals with respect to the predicted value is due to the underlying trend in the data over the observation period.

For comparison to standard linear prediction, we find the BLUP based on the estimated covariance matrix, and create Gaussian-based 95\% confidence intervals from the estimated MSPE.
When done on the standardized data, we obtain a coverage rate of 0.87, and when done on square-root transformed data to account for the skewness, we obtain a coverage rate of 0.66.

We also compare our prediction method to the extremes-based method of \cite{cooley2012approximating}, which approximated the conditional distribution of the large values of a regularly varying variate via a parametric model for the angular measure. 
The method of \cite{cooley2012approximating} can be applied due to this application's relatively low dimension.
The pairwise beta model \citep{cooley2010pairwise} is fit by maximum likelihood to the preprocessed training data set.
The 95\% prediction intervals are based on the approximated conditional density of $X_5$ given $x_1, \ldots, x_4$, and the achieved coverage rate for the test set is 0.965.
While the fitted angular measure model would seemingly contain more information than the estimated TPDM, we found that the widths of the prediction intervals were very similar for the two methods.  
The average ratio of the interval width from the parametric approach by \cite{cooley2012approximating} to our TPDM-based approach was 1.04.

We then apply our prediction method to five dates in 2019 and 2020 (including January 23, 2020 in Figure \ref{fig: washingtonDC}) when observed values at the four recording stations exhibit large radial components, with no observation taken at Alexandria.
Here, we use the full dataset from the period 1995-2014 to estimate the TPDM, and we obtain a slightly different estimate $\hat{\bm b}^{*} = (0.086, 0.204, 0.080, 0.465)^\top$ and the prediction TPDM estimate has an off-diagonal element of 0.559.
The right panel of Figure \ref{fig:no2} shows the point estimate and 95\% prediction intervals from our transformed linear approach, after back transformation to the original scale.
The trend at Arlington was used for the unobserved trend at Alexandria.
For comparison, covariance matrix-based BLUP estimates and MSPE-based 95\% prediction intervals for these dates are shown with the dashed line.
As expected, these intervals are narrower than those produced by our transformed-linear approach.

\subsection{UK precipitation}

We apply the transformed linear prediction method to a higher-dimensional precipitation data set from the Cumbria and Lancashire region in the UK.
The data, obtained from \cite{https://doi.org/10.5285/8ddfd4dd5af443f9ad382cd77366d877}, consist of daily rainfall records collected by Met Office observation stations across the UK.
We analyze data from 1960-2024, a period during which no clear long-term trends are observed.
The raw data contain missing values.
To ensure consistency across stations, we retain only the $30$ gauge stations for which the proportion of missing values is below $0.1$.
We further restrict the data set to days on which observations are available at all retained stations, resulting in $10,520$ observations.

Across these stations, maximum recorded rainfall ranges between 49.4 mm and 201.2 mm.
To assess marginal tail behavior, we estimate the shape parameter at each station by fitting a generalized Pareto distribution.
The average estimated shape parameter is $0.073$, with 80\% of the stations having positive estimates.
Although the data appear to be heavy-tailed, a marginal transformation is still needed so that we can assume $\alpha=2$.

Let $\bX^{(orig)}$ denote the random vector of daily rainfall.
For simplicity, we use a rank-based marginal transformation $X_i=(1-\widehat{F}_i(X_i^{(orig)}))^{-1/2}-\delta$, applied to each gauge station's data so that $X_i$ follows the same shifted Pareto distribution as in the air pollution application.
We again assume $\bX_t$, the random vector denoting the observations on day $t$, are i.i.d. copies of $\bX\in\Vq_+$.
We randomly select two-thirds of the data as a training set ($n_{train}=7,014)$ and use it to estimate the TPDM and obtain the vector $\hat{\bm{b}}^*.$
The remaining one-third of the data ($n_{test}=3,506$) serves as a test set to evaluate coverage rates.

\begin{figure}[th]
\centering
\includegraphics[width=0.4\textwidth]{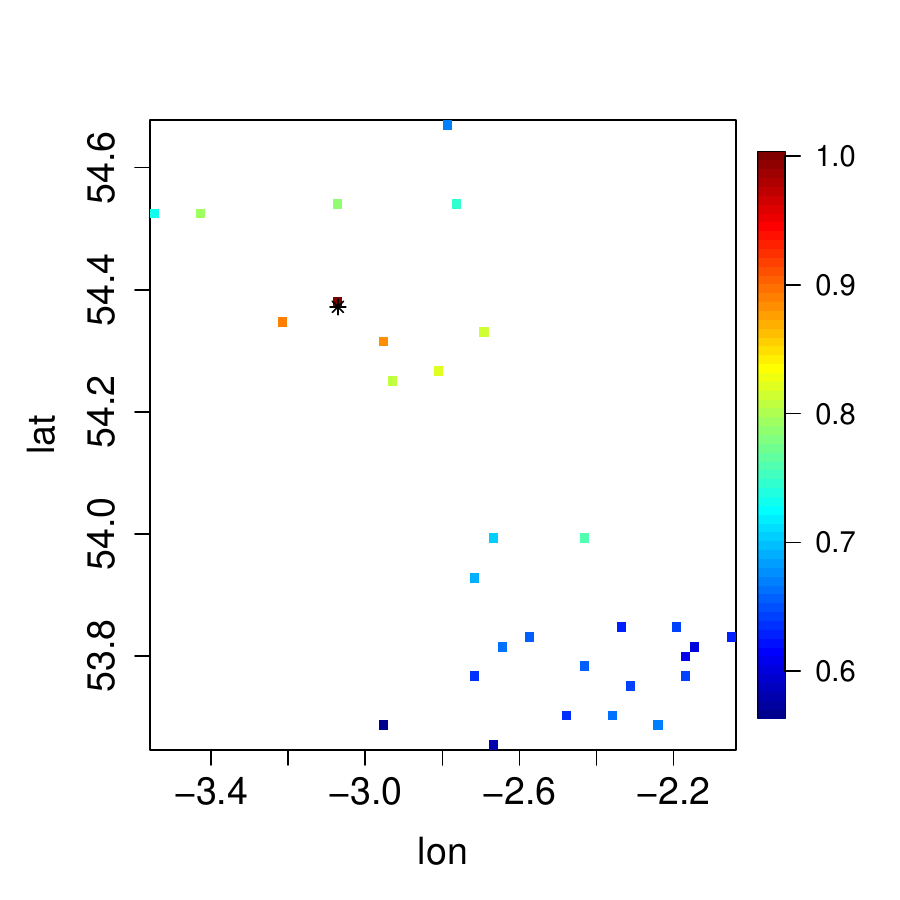}
\includegraphics[width=0.4\textwidth]{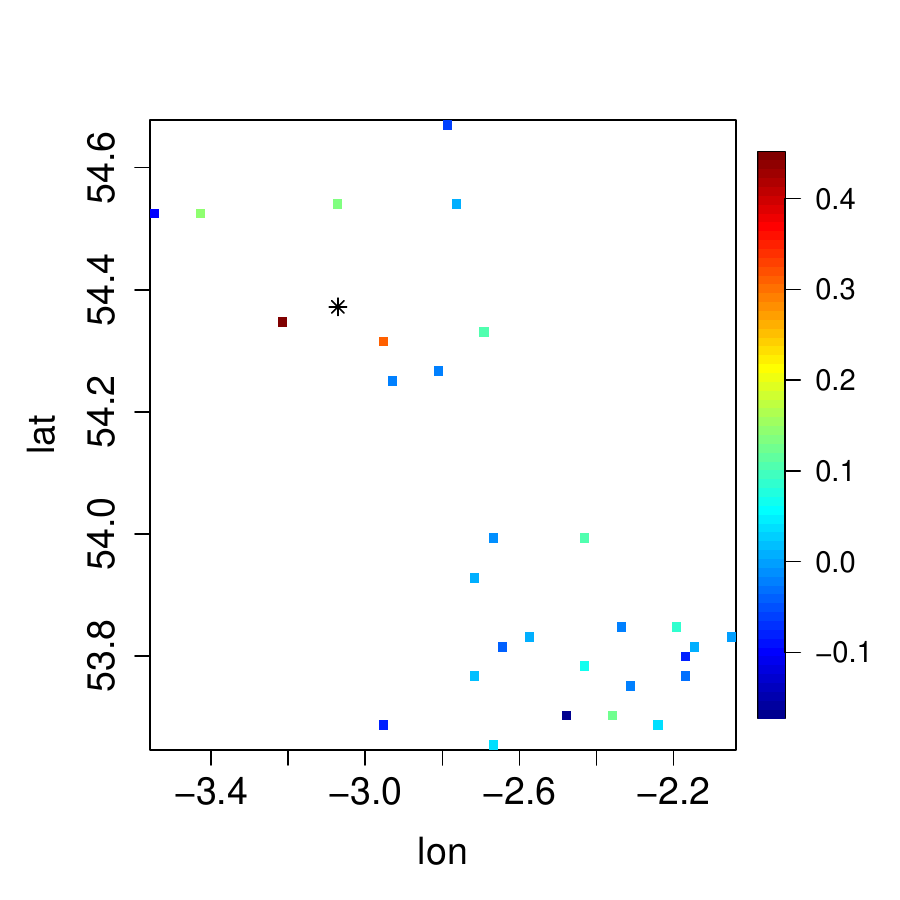}
\caption{(Left) Estimated TPDM estimates between the target station (marked with a star) and all remaining stations. (Right) Estimated weights based on the estimated TPDM.}
\label{fig:prcp}
\end{figure}

Following similar steps as in the previous application, we first estimate the $30\times 30$ TPDM $\Sigma_{\bX}$ from the training set using the 95\% quantile as the radial threshold.
We select station ID 12722 (Coniston-Garden House, Cumbria) as the prediction target, since it has the smallest proportion of missing values among the stations considered.
The left panel of Figure~\ref{fig:prcp} displays the pairwise extremal dependence between this target station and the remaining stations across the Cumbria and Lancashire regions, revealing stronger extremal dependence at nearby locations and weaker dependence with increasing distance.
The TPDM estimates show little sensitivity to the choice of threshold, and similar dependence patterns are observed across different target stations.

The right panel of Figure~\ref{fig:prcp} displays the estimated weights $\hat{\bm b}^*$, with larger magnitudes concentrated near the target station (marked with a star), reflecting the strongly localized nature of the extreme event.
As in the NO$_2$ application, we compute predictions $\hX_{t,30}$ for all $t$, and restrict attention to cases in which $\hX_{t,30}$ exceeds its 0.95 quantile.
Using the normalized angular measure obtained from the decomposition, along with the corresponding kernel density estimate, we evaluate the coverage of the transformed linear $95\%$ prediction intervals.
Using the default kernel density bandwidth returned by the \texttt{kde} method results in a coverage rate of $84\%$,
which we believe provides a reasonable representation of uncertainty in predicting extreme precipitation, despite being somewhat below the target nominal coverage.

Because we work in the imputation setting for this application; that is, the training set contains observations of the value $X_{p+1}$ which will be predicted in the test set, $k$-fold cross validation can be used to set a kernel density bandwidth which directly optimizes the coverage bandwidth.
We split the data into $k$-folds and, for each training set, estimate the TPDM and construct prediction intervals.
For each bandwidth in $\{0.1,0.15,\ldots,0.7\},$ kernel density estimation is carried out, and the test set is used to evaluate the coverage rate for a target coverage rate (e.g., 0.95).
The average coverage rate across splits is then computed, and the bandwidth yielding an average coverage closest to the target level is selected.
The resulting $5$-fold cross-validated bandwidth results in a coverage rate of 94.5\%.

For comparison with standard linear prediction, we compute the BLUP using the covariance matrix estimated from the training data and then obtain predictions for all $t$.
Gaussian-based 95\% confidence intervals are then constructed using the estimated MSPE, and evaluation is restricted to predictions for which predicted values exceed its 0.95 quantile.
After applying weights to square-root transformed data to account for skewness, the corresponding coverage rate is 0.31.


\addtolength{\textheight}{-.2in}%

\section{Summary and discussion}
\label{sec:summary}

We propose a method for linear prediction when observations are large.
Our approach constructs an inner product space of nonnegative random variables derived from transformed linear combinations of independent regularly varying random variables, which enables the derivation of the optimal transformed linear predictor via the projection theorem.
We illustrate its effectiveness for creating prediction intervals through both a simulation study and two data applications.
This linear method is straightforward, relying only on the TPDM, which is estimable even in high dimensions.

We have restricted our attention to nonnegative regularly varying random variables to focus on the upper tail, which is particularly relevant for environmental data.
Relaxing this restriction could allow one to use standard linear operations.
However, even for data with negative values, there is value in focusing in one direction.

The random vectors $\bm X = A \circ \bm Z$, constructed from our vector space, yields a straightforward angular measure form consisting of $q$ point masses, where $q$ is the number of columns of $A$.
While models with angular measures consisting of discrete point masses can be restrictive, our approach differs in that we do not need to specify $q$ to perform predictions.
Instead, by employing the information in the TPDM, we can easily obtain point predictions and construct prediction intervals that reflect the available information.

In many applications, direct dependence cannot be measured between the observed values and the value to be predicted.  
In kriging, for example, a spatial process model is first fit so that covariance between any two locations is quantified.
Similarly, one could model extremal pairwise dependence as a function of distance before applying the methods here to perform prediction at extreme levels.

\section*{Acknowledgments}
J.L. was partially supported by UK Engineering and Physical Sciences Research Council (EPSRC) grant EP/X010449/1 and 
D.C. was partially supported by US National Science Foundation Grant DMS-1811657.

\section*{Disclosure statement}\label{disclosure-statement}
The authors report there are no competing interests to declare.

\section*{Supplementary material}
\textbf{PDF Supplementary:} Contains theoretical proofs and additional numerical results from simulations.

\textbf{Datasets and Codes:} The original/pre-processed datasets, codes for implementing linear prediction in the simulation studies and applications, and the details needed to reproduce the results in Sections~\ref{sec:simulation} and~\ref{sec:applications} are available in the Github repository: \texttt{https://github.com/JeongjinLee88/extlinpred}.


\bibliography{bibliography.bib}

\end{document}